\newcommand{\bea}{\begin{eqnarray}}
\newcommand{\eea}{\end{eqnarray}}

\documentclass[superscriptaddress]{PoS}

\usepackage{graphicx}

\title{$\boldmath B \rightarrow D^* \ell \nu$ with 2+1 flavors}

\ShortTitle{$B \rightarrow D^* \ell \nu$ with 2+1 flavors}

\author{\speaker{Jack Laiho}%
        \\Theoretical Physics Department,
        Fermi National Accelerator Laboratory\thanks{Operated by Fermi  Research Alliance, LLC,
         under Contract No.~DE-AC02-07CH11359 with the  United States Department of Energy.},
        Batavia, Illinois, USA \\
        Physics Department,
        Washington University,\thanks{present address}
        St.\ Louis, Missouri, USA \\
        E-mail: \email{jlaiho@fnal.gov}}
\author{Fermilab Lattice and MILC Collaborations}

\abstract{We present a calculation of the form factor for
$B\rightarrow D^* l \nu$ using a 2+1 improved staggered action for
the light quarks (on the MILC configurations), and the Fermilab
action for the heavy quarks.  The form factor is computed at zero
recoil using a new double ratio method which yields the form
factor more directly than previous approaches.}

\FullConference{The XXV International Symposium on Lattice Field Theory\\
         July 30 - August 4 2007\\
         Regensburg, Germany}

\begin{document}

\section{Introduction}

The CKM element $V_{cb}$ is important for the phenomenology of
flavor physics in determining the apex of the unitarity triangle
in the complex plane.  For example, the Standard Model prediction
of $\epsilon_K$ depends sensitively on $V_{cb}$ (where it appears
to the fourth power), and the present errors on this quantity
contribute errors to $\epsilon_K$ of the same size as those due to
$B_K$, the kaon mixing parameter which has been the focus of much
recent work \cite{gamiz, cohen, vandewater}. It is possible to
determine $|V_{cb}|$ from both inclusive and exclusive
semileptonic $B$ decays, and they are both limited by theoretical
uncertainties. The inclusive method makes use of the heavy quark
expansion \cite{ball, bigi}, but is limited by the breakdown of
local quark-hadron duality, the errors of which are difficult to
estimate.  The exclusive method requires reducing the uncertainty
of the form factor ${\cal F}_{B\rightarrow D^*}$, which has been
calculated using lattice QCD in the quenched approximation
\cite{hashimoto}.  Given the phenomenological importance of this
quantity we have revisited this calculation of ${\cal
F}_{B\rightarrow D^*}$ using the 2+1 flavor MILC lattices with
improved light staggered quarks \cite{MILC}. The quenching error
is thus eliminated, and the systematic error associated with the
chiral extrapolation is reduced significantly.

This calculation was done using a blind analysis as follows: the
perturbation theory calculation needed to renormalize the lattice
current was done separately from the rest of the numerical
analysis, and the renormalization constants needed to compare
results at different lattice spacings to the continuum were given
an overall offset which was not revealed until the systematic
errors in the rest of the numerical analysis had been determined.

\section{Obtaining $|V_{cb}|$}

The differential rate for the semileptonic decay $\overline{B}\to
D^*l\overline{\nu}_l$ is

\bea \frac{d\Gamma}{dw} &=&
\frac{G^2_F}{4\pi^3}m^3_{D^*}(m_B-m_{D^*})^2\sqrt{w^2-1}{\cal
G}(w)|V_{cb}|^2|{\cal F}_{B\rightarrow D^*}(w)|^2 \eea

\noindent where $w=v' \cdot v$ is the velocity transfer from the
initial state to the final state, and ${\cal G}(w)|{\cal F}_{B
\rightarrow D^*}|^2$ contains a combination of four form factors
which must be calculated nonperturbatively. At zero recoil ${\cal
G}(1)=1$, and ${\cal F}_{B\rightarrow D^*}(1)$ reduces to a single
form factor, $h_{A_1}(1)$.  This is sufficient to determine
$|V_{cb}|$ from experiment. Heavy quark symmetry plays an
important role in constraining $h_{A_1}(1)$, leading to the heavy
quark expansion \cite{falk, mannel}

\bea h_{A_1}(1) &=&
\eta_A\left[1-\frac{\ell_V}{(2m_c)^2}+\frac{2\ell_A}{2m_c
2m_b}-\frac{\ell_P}{(2m_b)^2}\right], \nonumber \\ &&
\label{eq:hA1}\eea

\noindent up to order $1/m_Q^2$ and where $\eta_A$ is a factor
which matches QCD and heavy quark effective theory (HQET).  The
$\ell$'s are long-distance matrix elements of the heavy quark
effective theory. The earlier work by the Fermilab lattice
collaboration \cite{hashimoto} used a series of three double
ratios in order to obtain separately each of the three $1/m_Q^2$
coefficients in Eq.~(\ref{eq:hA1}). These three double ratios also
determine three out of the four coefficients appearing at
$1/m_Q^3$ in the heavy quark expansion.  It was shown in
\cite{kronfeld} that for the Fermilab method matched to tree level
in $\alpha_s$ and to leading order in HQET, the leading
discretization errors for the double ratios for this quantity are
of order $\alpha_s(\overline{\Lambda}/m_Q)^2$ and
$\overline{\Lambda}/m_Q^3$.

In the calculation reported here, the form factor $h_{A_1}(1)$ is
computed more directly using only one double ratio,

\bea\label{eq:doubleR} {\cal R}_{A_1}=\frac{\langle
D^*|\overline{c}\gamma_j \gamma_5 b|\overline{B}\rangle \langle
\overline{B}|\overline{b}\gamma_j \gamma_5 c|D^*\rangle}{\langle
D^*|\overline{c}\gamma_4 c|D^*\rangle \langle
\overline{B}|\overline{b}\gamma_4 b|\overline{B}\rangle} =
\left|h_{A_1}(1)\right|^2. \eea

\noindent which is exact to all orders in the heavy quark
expansion (modulo discretization errors for the corresponding
lattice ratio). The errors in this ratio do not rigorously scale
as ${\cal R}-1$ because Eq.~(\ref{eq:doubleR}) is not one in the
limit of equal bottom and charm quark masses (it becomes one only
in the static quark limit).  Nevertheless, this double ratio still
retains the desirable features of the previous double ratios, i.e.
large statistical error cancellations, and the cancellation of
most of the lattice current renormalization.  The quenching error
has been eliminated by including the fermion determinant in the
weighting of the gauge configurations, and so the rigorous scaling
of all the errors as ${\cal R}-1$, including the quenching error,
is no longer as important.  The more direct method introduced here
has the significant advantage that extracting coefficients from
fits to HQET expressions as a function of heavy quark masses is
not necessary, and no error is introduced from truncating the
heavy quark expansion to a fixed order in $1/m_Q^n$.

Most of the current renormalization cancels in the lattice double
ratio, leaving only a small correction factor, $\rho$, defined
such that $\rho \sqrt{R_{lat}} = \sqrt{{\cal R}_{cont}} = h(1)$,
as discussed in \cite{harada}. This $\rho$ factor has been
calculated perturbatively \cite{elkhadra}, and was found to
contribute less than a $0.5\%$ correction.

\section{Lattice calculation}

The lattice calculation was done on the MILC lattices for three
lattice spacings ($a \approx 0.15$, $0.125$, and $0.09$ fm) where
the light quarks were computed with the ``AsqTad'' staggered
action. The heavy quarks were computed using the clover action
with the Fermilab interpretation in terms of HQET
\cite{elkhadra2}. We have several light masses at both full QCD
and partially quenched points ($m_{valence}\neq m_{sea}$), and our
light quark masses range between $m_{s}/10$ and $m_s/2$.

\begin{figure}\label{fig:doubleR}
\bigskip
\bigskip
\begin{center}
\includegraphics[scale=.42]{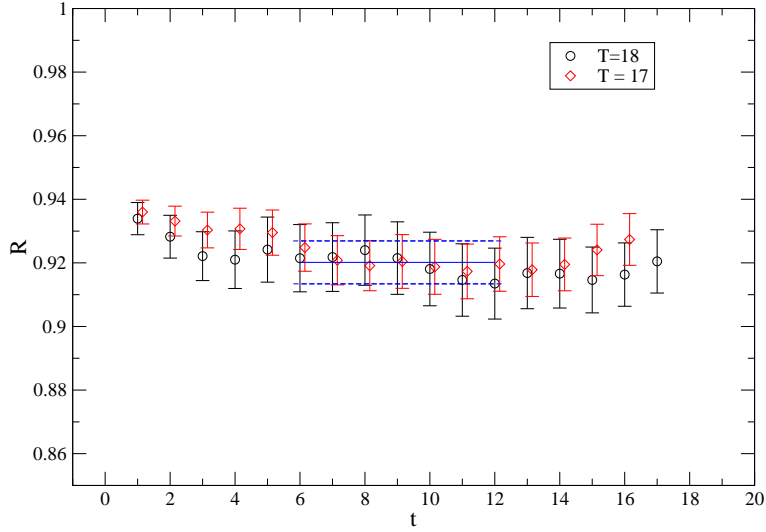}
\end{center}
\caption{Double ratio on the $m_{\ell}= 0.0124$ fine ensemble. The
source was fixed to time slice 0, and the operator position was
varied as a function of time.  Two different sink (extension)
points were used with even and odd time separations between source
and sink [$C(0,t,T)$ and $C(0,t,T+1)$] in order to study the
effect of non-oscillating wrong parity states.  The fit is to the
average of the source sink separations given in Eq.~(3.1).}
\end{figure}

Extracting correlation functions that contain staggered quarks
presents an extra complication because of the contributions of
wrong parity excited states which introduce oscillations into the
usual plateau fits. The average,

\bea\label{eq:avg} C^{X\to Y}_{avg}(0,t,T) &\equiv &
\frac{1}{2}C^{X\to Y}(0,t,T) + \frac{1}{4}C^{X\to Y}(0,t,T+1) +
\frac{1}{4}C^{X\to Y}(0,t+1,T+1),\eea

\noindent is equivalent to a smearing which suppresses the
oscillating states, and has been applied to all of the data for
the double ratios.  Figure~(\ref{fig:doubleR}) shows a plateau fit
to the double ratio used to obtain $h_{A_1}(1)$.  The source is at
time slice 0, the sink is at $T$, and the operator position is
varied along $t$. Two different extended propagators were
constructed at even and odd source sink separations ($T=16,17$).
The average of these two extension points was taken according to
Eq.~(\ref{eq:avg}), and this average was fit to a constant as
shown in Figure~1.  There is no detectable oscillation even before
the average is taken; the oscillating contributions are reduced
even further in the average so that their systematic errors can be
safely neglected.

The chiral extrapolation errors can be controlled by using the
appropriate rooted staggered chiral perturbation theory
(rS$\chi$PT) for heavy light quantities \cite{aubin}. Eq.~(34) of
\cite{laiho} gives the expression needed for fits to $h_{A_1}(1)$
for partially quenched data with degenerate up and down quark
masses (the 2+1 case).
This partially quenched expression parameterizes the dependence on
both valence and sea quark masses, and includes taste breaking
violations coming from the light quark sector.  The expression
contains explicit dependence on the lattice spacing $a$, and
requires as inputs the parameters of the staggered chiral
lagrangian $\delta'_V$, $\delta'_A$, in addition to the staggered
taste splittings $\Delta_{P,A,T,V,I}$.  These parameters can be
obtained from chiral fits to the pseudoscalar sector and are held
fixed in the chiral extrapolation of $h_{A_1}(1)$. The continuum
low energy constant $g_{D^*D\pi}$ appears, and this can be taken
from phenomenology; we take a generous range of values for this
term to estimate the error it contributes to $h_{A_1}(1)$. The
only other parameter which appears at NLO is an overall constant
that is determined by a fit to our data for $h_{A_1}(1)$.

For the chiral fits we find it useful to form two ratios that
normalize results for $h_{A_1}(1)$ at a ``fiducial point,''

\bea R_{sea}(m_L, m_S, a) = \frac{h_{A_1}(m_x^{\rm fid}, m_L, m_S,
a)}{h_{A_1}(m_x^{\rm fid}, m_L^{\rm fid}, m_S^{\rm fid}, a)},
 \ \ \ \ R_{val}(m_L, m_S, a) = \frac{h_{A_1}(m_x, m_L, m_S,
a)}{h_{A_1}(m_x^{\rm fid}, m_L, m_S, a)}. \eea

\noindent where fid stands for fiducial, $m_x$ is the light
valence quark, $m_L$ is the light sea quark, $m_S$ is the strange
sea quark.  Here we take $m_x^{\rm fid} \approx 0.5 m_{\rm
strange}^{\rm physical}$, $m_L^{\rm fid} \approx 0.5 m_{\rm
strange}^{\rm physical}$, and $m_S^{\rm fid} \approx m_{\rm
strange}^{\rm physical}$. The ratios in Eq.~(3.2) are now
quadruple ratios; thus the statistical errors and excited state
contamination are further suppressed over that of the double
ratio. The main advantage of these ratios, however, is that heavy
quark discretization effects largely cancel, so that we can
disentangle the heavy quark discretization effects and those of
the staggered chiral logs. This isolates the discretization
effects coming from non-analytic taste violations, and these can
be removed using rS$\chi$PT. We have chosen the fiducial point to
be $\approx 0.5 m_{\rm strange}^{\rm physical}$ because it would
be feasible to simulate this mass point on very fine lattices and
smaller volumes without running into finite size effects, thus
normalizing our data at a point where the heavy quark
discretization effects are much smaller. For now we use the point
with $m \approx 0.5 m_{\rm strange}^{\rm physical}$ on the finest
lattice spacing available ($a\approx 0.09$ fm) as our fiducial
point.  By taking the chiral extrapolation and the continuum limit
of the two ratios, multiplying them together and then multiplying
that by the value of $h_{A_1}(1)$ at the fiducial mass on the
finest available lattice spacing, we can construct the value of
the form factor at the physical light quark mass,
$h_{A_1}^{phys}=h_{A_1}^{fid}\times [R_{sea}(m^{phys}_{\ell},
m_s^{phys}, 0)\times R_{val}(m^{phys}_{\ell}, m_s^{phys}, 0)]$.
This quantity is shown in Figure~2.

\begin{figure}\label{fig:chiralFullQCD}
\bigskip
\bigskip
\begin{center}
\includegraphics[scale=.42]{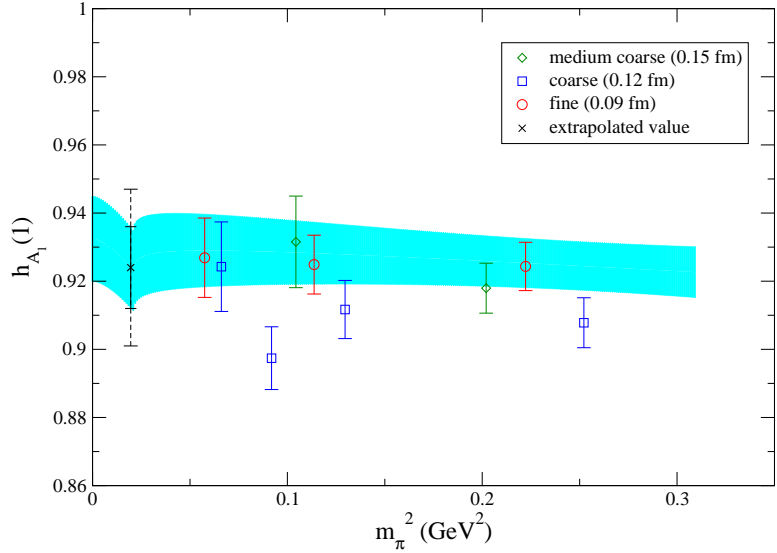}
\caption{All of the data at the full QCD points
($m_{valence}=m_{sea}$) on the three lattice spacings. The cyan
(light grey) band is the continuum extrapolated full QCD curve.
The cross is the value at the physical light pion mass, where the
solid line is the statistical error, and the dashed line is the
total systematic error added to the statistical error in
quadrature.}
\end{center}
\end{figure}

\begin{figure}\label{fig:aSq}
\bigskip
\begin{center}
\includegraphics[scale=.42]{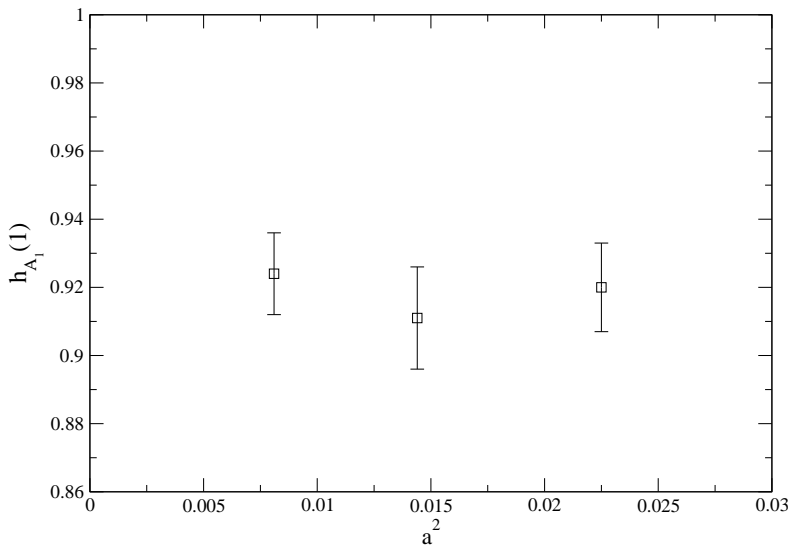}
\caption{The values for $h_{A_1}$ using the continuum extrapolated
ratios determined in the fiducial point procedure to extrapolate
the fiducial points on each of the three lattice spacings to the
physical light quark masses.  The fiducial point procedure allows
us to remove the taste violations coming from staggered chiral
logs, but it does not remove the analytic terms associated with
the light quark sector, nor does it remove the heavy quark
discretization errors.   Although it is appropriate to extrapolate
this curve to the continuum, a first principles extrapolation
formula is not known.  We therefore compare the value of $h_{A_1}$
using the fiducial point on the fine lattice with the results
obtained by using fiducial points on coarser lattice spacings. A
comparison of the scatter of these results allows us to estimate
the size of the remaining light quark and heavy quark
discretization errors.}
\end{center}
\end{figure}

\section{Results and conclusions}

The final error budget is presented in Table 1.  The error
labelled ``$g_{D^*D\pi}$ uncertainty'' comes from the error in the
chiral low energy constant $g_{D^* D \pi}$, which we take to vary
between 0.3 and 0.6.  The next error is the difference between
doing NLO chiral fits for the chiral extrapolation, versus fits
which include the NNLO analytic terms but not the 2-loop
logarithmic terms, which have not been calculated. Both fits give
acceptable confidence levels.

Our largest systematic uncertainty comes from discretization
errors.  The fiducial point procedure described above allows us to
remove the effect of the splittings in the staggered chiral logs,
but it does not determine and remove the analytic $a^2$ dependence
in the light quark sector, nor the heavy quark discretization
errors. Comparing the values obtained with different fiducial
points on various lattice spacings gives an estimate of the size
of the remaining light quark and heavy quark discretization
errors.  The scatter of the points in Figure 2 gives an estimate
of the size of these effects, which cannot be resolved within
statistics.  The difference between the fine ($a=0.09$ fm) and
coarse ($a=0.12$ fm) lattice spacings is a $1.3\%$ difference,
which is about the size one would expect for heavy quark
discretization errors in this quantity from power counting
arguments and a reasonable choice for the HQET parameter
$\overline{\Lambda}$.

The error labelled ``kappa tuning'' comes from the parametric
uncertainty associated with tuning the charm and bottom quark
masses.  The next error is from the perturbative matching of the
lattice currents in the double ratio.  As mentioned above, this
renormalization factor is small because most of the
renormalization cancels nonperturbatively in the ratio.  We take
the entire 1-loop correction of $0.4\%$ as a conservative estimate
of the error due to the omission of higher orders.

We quote a preliminary result for the form factor
$h_{A_1}(1)=0.924(12)(19)$, where the first error is statistical,
and the second is the sum of all systematic errors in quadrature.
Taking the latest world average of ${\cal F}(1)|V_{cb}|=(36.0 \pm
0.6)\times 10^{-3}$ from experiment \cite{hfag}, we find
$|V_{cb}|=(38.7 \pm 0.7_{exp} \pm 0.9_{theo})\times 10^{-3}$.  We
estimate that the theoretical error on this determination of
$|V_{cb}|$ from exclusive $B\rightarrow D^* \ell \nu$ can be
reduced significantly by making use of the existing extra-fine
MILC lattices ($a=0.06$ fm) and higher statistics on the coarser
ensembles.

\begin{table}
\begin{center}
\caption{Error budget}
 \label{tabtwo}
\begin{tabular}{cc}
\hline \hline
  uncertainty & $h_{A_1}(1)$  \\
  \hline
  statistics &  $1.2\%$   \\
  $g_{D^*D\pi}$ uncertainty &  $0.6\%$  \\
  NLO vs partial NNLO ChPT fits &  $0.9\%$  \\
  discretization errors &  $1.3\%$  \\
  kappa tuning &  $1.0\%$  \\
  perturbation theory & $0.4\%$  \\
\hline \hline \\
   Total & $2.3\%$ \\
\end{tabular}
\end{center}
\end{table}

\end{document}